\documentclass{article}
\usepackage[dvips]{graphicx}
\usepackage{rotating}
\begin{document}
\bibliographystyle{plain}

\begin{center}
\Large Another Monte Carlo Renormalization Group Algorithm \\

\normalsize John P. Donohue\\
Department of Physics, University of California,Santa Cruz \\
Santa Cruz, CA 95064 \\
\date{\today}

\end{center}

%\maketitle
%\pagestyle{empty}

\begin{abstract}
A Monte Carlo Renormalization Group algorithm is used on the Ising 
model to derive critical exponents and the critical temperature. 
The algorithm is based on a minimum relative entropy 
 iteration developed previously to derive potentials from 
equilibrium configurations. This previous algorithm is modified
to derive useful information in an RG iteration.
The method is applied in several dimensions with limited success.
Required accuracy has not been achieved, but the method is interesting.
\end{abstract}

\section*{Introduction}
Monte Carlo Renormalization Group has been used with much success for
over twenty-five years. 
The Large Cell Renormalization method calculates the 
renormalized parameters by renormalizing the system to a two spin system
which can be solved exactly.
Swendsen's method avoids calculating the renormailized parameters by 
using a Taylor expansion of the Renormalization operator. 
Calculating several variances leads to a determination of
the eigenvalues of the operator \cite{Swendsen,mcmisp,landaubinder}.
Ma's method calculates the parameters in the Renormalized 
Hamiltonian using an analysis of the Monte Carlo 
dynamics \cite{landaubinder,ma}. 
In our method, the
parameters of the Renormalized Hamiltonian are derived using a
minimum relative entropy iteration.
The number of coupling parameters is controlled and error
in the method can be estimated using usual statistical methods.
The eigenvalues of the operator can be determined directly from
the Renormalized Hamiltonian. This also allows the critical temperature
to be directly calculated.
Higher order coupling parameters can be included for an estimate of
their effect.

\section*{Our Method}

The following is based largely on an algorithm originally developed by 
reference \cite{det2}.
The method is based on the following assumptions. The system is assumed 
to be in thermodynamic equilibrium and 
the energy can be written as a sum of terms which are products
of parameters and functions of the configuration.
$ E(\Gamma,\vec{P}) = \sum_{i} p_{i} * h_{i}(\Gamma) = \vec{P} \cdot \vec{H}$
where $\Gamma$ represents the configuration of the system(s).
$\vec{P} = \{p_{i}\}$ represents the set of parameters to be derived.
In the case of the Ising Model, 
$ E = -J \sum_{<ij>} s_i \cdot s_j - K \sum_i s_i $. 
In our notation, $h_1 = \sum_{<ij>} s_i \cdot s_j$ , $p_1= -J$ , 
$ h_2 = \sum_i s_i$ and $p_2= -K $.

The probability of a configuration, given parameters, is given by the 
Boltzmann distribution
$ Prob(\Gamma | \vec{P}) = e^{-E(\Gamma , \vec{P})/kT} / Z = e^{(-E(\Gamma , \vec{P}) + F(\vec{P})) / kT} $ ,where
$ Z(\vec{P}) = \sum_{\Gamma}Exp(-\beta E(\Gamma,\vec{P}))$ and
$ F(\vec{P}) =  -kT \ln(Z(\vec{P}))$.

If we are given the exact equilibrium conformation, $\Gamma^*$ ,
the maximum likelihood of parameter values are those values for which the 
probability, $Prob(\Gamma^* | \vec{P})$ is a maximum wrt $\vec{P}$.
Maximizing an exponential corresponds
to maximizing the argument (ignoring the multiplicative constant $\beta$), 
$ -E(\Gamma^*,\vec{P}) + F(\vec{P}) = Q(\vec{P}) $.
This also corresponds to extremizing the entropy $TS = E - F $.

Our method is basically the multi-dimensional form of 
Newton's method for optimizing functions.
Maximizing $Q(\vec{P})$, Newton's Method is
\begin{equation}
\vec{P}^{k+1} = \vec{P}^{k} - D^2(Q(\vec{P}^{k}))^{-1} \cdot D(Q(\vec{P}^{k}))
\end{equation}
where $(D^2)^{-1}$ represents the inverse Hessian matrix 
and D represents the gradient.
In practice this is modified slightly,
\begin{equation}
\vec{P}^{k+1} = \vec{P}^{k} + \epsilon(\Delta \vec{P})
\end{equation}
where the use of $ \epsilon < 1 $ corresponds to the "Damped Newton's Method".

Maximizing $ Q = -E + F $ wrt $\vec{P}$
using statistical mechanical definitions leads to the following 
\begin{equation} D(Q)_i = -h_i^* + <h_i> \end{equation} 
\begin{equation} D^2(Q)_{i,j} = \beta ( <h_i> <h_j> - <h_i h_j> ) = - \beta Cov(h_i,h_j) \end{equation}
Resulting in the following iterative
equation where $VCM(\vec{H})$ is the variance-covariance matrix of $\vec{H}$
\begin{equation} \Delta \vec{P} = kT*VCM(\vec{H})^{-1} \cdot (<\vec{H}> - \vec{H^*} )\end{equation}
The method is easily generalized to a distribution of equilibrium configurations.
\begin{equation} \Delta \vec{P} =  kT*VCM(\vec{H})^{-1} \bullet (<\vec{H}> - <\vec{H}>_{Prob(\Gamma)}) \end{equation}
$<...>$ represents a Boltzmann average and $<...>_{Prob(\Gamma)}$ 
represents an average over the given distribution.

\section*{Renormalization}
Renormalization can be described as an operator acting 
on the original Hamiltonian to create a Renormalized Hamiltonian.
Linearizing the operator near the fixed point, relevent 
eigenvalues can be isolated.
By examining the plot of renormalized vs original parameters, 
the eigenvalues can be determined. 
The eigenvalues $\lambda_i$, correspond to the slope of these lines.
With the usual approximations, the exponents
can be calculated , $\nu_i = \ln(b)/\ln(\lambda_i) $ \cite{landaubinder}.
The critical temperature can be determined by calculating the intersection
of the derived thermal coupling parameters and the line y=x.
Flow diagams could easily be extracted, although they are not done in this
study.

Using Renormalization notation, the reduced hamiltonian for
the Ising model is given below.
\begin{equation} H_r = - H / kT \end{equation}
\begin{equation}  H_r = K_0\Sigma s_i +  K_1 \Sigma_{nn}s_i s_j + K_2 \Sigma_{nnn}s_i s_j + K_4\Sigma_{square}s_i s_j s_k s_l + ... \end{equation}
Applying the algorithm to Renormalization requires an outer iteration
over a parameter (ie $K_i$ for some i) while other parameters remain fixed.

The original system(s) is brought to equilibrium at some fixed values
for all parameters. 
The renormalized system(s) is created from this original system.
The above iteration represented by equations 1,2 and 5 can be carried out
on large number of lattices simultaneously 
to determine what values of parameters
would lead to this renormalized system. 
As a parameter is varied
 the corresponding eigenvalues and critical exponents can be derived.
The method is simpler to apply if the renormalization method is not a 
quasi-linear method \cite{thefish}. 
Majority rule renormalization is used in all of the following.

\section*{2-D Ising Model}
The algorithm was attempted on a 2-D Ising model. Exponents 
and critical temperature are
known from the original solution by Onsager \cite{onsager,cresw}.

$ \nu=1 $ , $  \theta=8/15 $ and $ T_{critical}=2.269 $

\subsection*{2D Ising Model Thermal Coupling Exponent}

The thermal eigenvalue can be determined
by varying the ratio of $J/kT$ through the critical value at zero field. 
The corresponding exponent, in terms of correlation length and reduced
temperature, is 
\begin{equation} \xi \propto |t|^{-\nu} \end{equation}
The renormalized
parameters were calculated as per the above algorithm.
Figure 1 shows several runs of various lattice sizes
and number. The slopes, exponents and critical temperatures are 
shown in table 1.

\begin{figure}[h]
\includegraphics[angle=-90,width=4in]{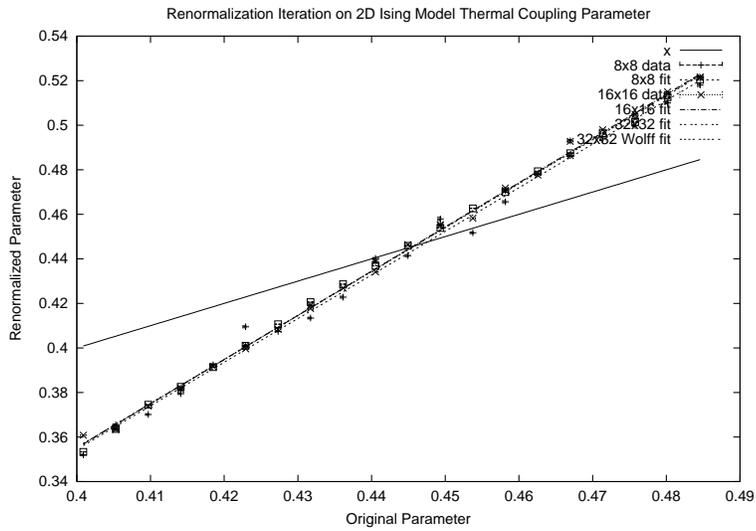}
\caption{Plot of Renormalized parameter $K_1'$ (Y-axis)
vs $K_1$ (X-axis) for 2-D Ising model.
Intersection of the line y=x with the data corresponds to the fixed point. 
Error bars are included on the plot but are on the scale of the size 
of the points.}
\end{figure}

\begin{table}[h]
\begin{tabular}{l|r|r|r|r}
\hline
Size & Slope & $\nu = \ln 2 / \ln(slope) $ & Intercept & $T_{critical}$ \\
\hline
8 & $ 1.96 \pm 0.04$ &  $ 1.03 \pm 0.03 $  & $ 0.43 \pm 0.02 $ & $2.23 \pm 0.14$ \\
\hline
16 & $ 1.99 \pm 0.02$ & $ 1.01 \pm 0.01$  & $ 0.44 \pm 0.01 $ & $2.25 \pm 0.07$ \\
\hline
32 & $ 2.02 \pm 0.03$ & $ 0.99 \pm 0.02$  & $ 0.45 \pm 0.01 $ & $2.27 \pm 0.08$ \\
\hline
32(W) & $ 1.95 \pm 0.01$ & $ 1.03 \pm 0.01$  & $ 0.42 \pm 0.01 $ & $2.26 \pm 0.06$ \\
\hline
64(W) & $ 1.98 \pm 0.02$ & $ 1.01 \pm 0.02$  & $ 0.43 \pm 0.01 $ & $2.28 \pm 0.07$ \\
\hline
\end{tabular}
\caption{Calculation of 2-D Ising model thermal coupling exponent.
Exact value is $\nu=1$ and $T_{critical}=2.269$}
\end{table}

\clearpage

\subsection*{Magnetic eigenvalue}

The other exponent and eigenvalue can be determined by fixing
 J/kT at approximately the critical value and varying
the field parameter through zero.
The corresponding exponent, in terms of correlation length and field,  is 
\begin{equation} \xi \propto |K|^{-\theta} \end{equation}
K represents the external field.
The renormalized parameters were calculated as per the above algorithm.
Figure 2 shows data runs over a large variation in field parameter.
The critical region, near zero field, shows a more linear 
relationship as shown in figure 3.
Least squares best fit to a straight line was done on this central region.
The results are shown in table 2.

\begin{figure}[h]
\includegraphics[angle=-90,width=4in]{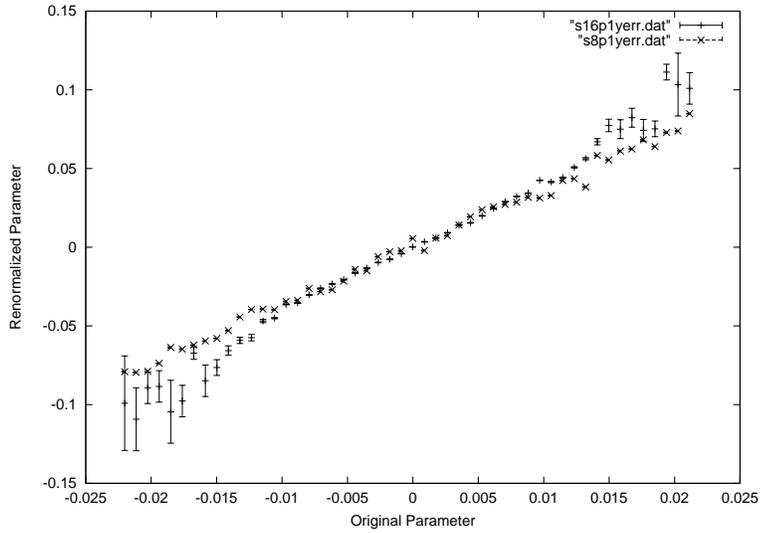}
\caption{Plot of Renormalized magnetic parameter/kT (Y-axis)
vs Original magnetic parameter/kT (X-axis) for 2-D Ising model 
near the critical temperature. 4000 MC steps per site were used.
Data is from systems with a.)16x16,n=1024 b.)8x8,n=1024.}
\end{figure}

\begin{figure}[h]
\includegraphics[angle=-90,width=4in]{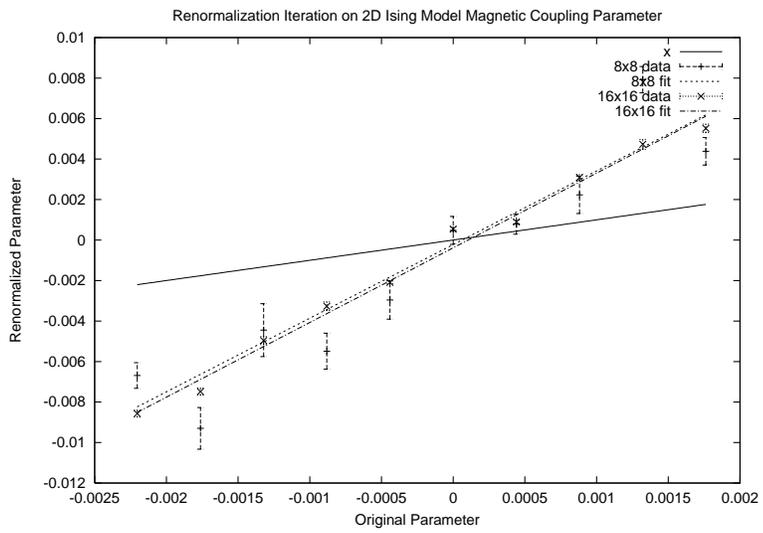}
\caption{Plot of Renormalized magnetic parameter/kT (Y-axis)
vs Original magnetic parameter/kT (X-axis) for 2-D Ising model 
near the critical temperature.}
\end{figure}

\begin{table}[h]
\begin{tabular}{l|r|r|r}
\hline
Size & Slope & $\theta = \ln 2 / \ln(\lambda)$  \\
\hline
8 & $3.63 \pm  0.48 $ & $ 0.54\pm 0.06$ \\
\hline
16 & $3.68 \pm  0.16$ & $0.53 \pm 0.02$\\
\hline
\end{tabular}
\caption{Calculation of magnetic exponent for 2-D Ising model.
Exact value is $\theta=8/15=0.53$}
\end{table}

\clearpage
\subsection*{3D Ising Model Eigenvalues}
The algorithm was applied to the 3D and 4D Ising model 
with encouraging results, even on small lattice sizes. 
Approximate exponents for 3D are known from several sources. 
Reviews are given by references \cite{cresw, landaubinder}. 
\begin{table}[h]
\begin{tabular}{l|r|r|r|r}
\hline
Size & Slope=a & $\nu = \ln 2 / \ln(slope)$ & Intercept=b & $T_{critical}= (1-a)/b $ \\
\hline
4 & $ 3.02 \pm 0.07$ &  $ 0.63 \pm 0.01 $  & $ 0.44 \pm 0.02 $ & $4.59 \pm 0.17 $ \\
\hline
32 & $ 3.40 \pm 0.09$ &  $ 0.57 \pm 0.01 $  & $ 0.53 \pm 0.02 $ & $4.53 \pm 0.24 $ \\
\hline
\end{tabular}
\caption{Calculation of thermal coupling exponent for 3-D Ising model.
Approximate value is $\nu \approx 0.63$ and 
$T_{critical} \approx 4.52$.}
\end{table}
\begin{table}[h]
\begin{tabular}{l|r|r}
\hline
Size & Slope & $\theta = \ln 2 / \ln(slope)$ \\
\hline
4 & $6.3 \pm  0.3 $ & $ 0.38 \pm 0.01$ \\
\hline
\end{tabular}
\caption{Calculation of magnetic exponent for 3-D Ising model. 
Expected value $\theta \approx 0.40 $ }
\end{table}
\clearpage
\subsection*{4D Ising Model - Mean Field Theory Exponents}
\begin{table}[h]
\begin{tabular}{l|r|r|r|r}
\hline
Size & Slope=a & $\nu =  \ln 2 / \ln(slope)$ & Intercept=b & $T_{critical}= (1-a)/b $ \\
\hline
4 & $ 4.3 \pm 0.1 $ &  $ 0.48 \pm 0.04 $  & $ 0.48 \pm 0.02 $ & $6.88 \pm 0.5 $ \\
\hline
\end{tabular}
\caption{Calculation of thermal coupling exponent for 4-D Ising model.
Exact values are $\nu = 0.5$ and $T_{critical} = 6.68 $. }
\end{table}

\begin{table}[h]
\begin{tabular}{l|r|r|r}
\hline
Size & Slope & $\theta $ & exact \\
\hline
4 & $ 10.2 \pm  0.8 $ & $ 0.30 \pm 0.01$ & 0.333.. \\
\hline
\end{tabular}
\caption{Calculation of magnetic exponent for 4-D Ising model. 
Exact value is $\theta = 1/3 $. }
\end{table}

\clearpage

\section*{Conclusion}
Much higher precision and accuracy is required to compare with
current estimates of exponents, but
the derived exponents match reasonably well with expected values.
Improvement is expected with larger lattices and more systems,
perhaps with more clever averaging methods.
The effect of critical slowing down has not been thouroughly 
investigated.
The algorithm has several advantages over other algorithms.
The number of coupling parameters is kept the same from
original system to the renormalized system with minimum
relative entropy derived values for the renormalized parameters.
This seems more consistent than an arbitrary cutoff in parameter space.
The renormalized parameters are directly calculated, unlike other algorithms.
By changing the parameter of the outer iteration, all exponents can
be magnified and derived.
If flow diagrams are desired, the algorithm could be easily modified
to provide them.
The error in the calculated exponents can be derived through 
usual statistical properties.

\section*{Acknowledgments}

J Deutsch, AP Young, Lik Wee Lee, Leif Poorman, Stefan Meyer, TJ Cox, B Allgood, D Doshay , Hierarchical Systems Research Foundation

\bibliography{mybib}

\end{document}